\documentclass[aps,prb,superscriptaddress,reprint,floatfix]{revtex4-1}
\usepackage{amsmath}
\usepackage{graphicx}
\usepackage{longtable}
\usepackage{colortbl}
\usepackage[english]{babel}
\usepackage{array}
\usepackage{booktabs}

\graphicspath{{graphs/}}

\begin{document}

\title{First-principles studies of the infrared spectra in liquid water from a systematically improved description of H-bond network}
\author{Jianhang Xu}
\affiliation{Department of Physics, Temple University, Philadelphia, PA 19122, USA}
\author{Mohan Chen}
\affiliation{CAPT, HEDPS, College of Engineering, Peking University, Beijing 100871, China}
\author{Cui Zhang}
\affiliation{Institute of Physics, Chinese Academy of Sciences, Beijing 100190, China}
\author{Xifan Wu}
\affiliation{Department of Physics, Temple University, Philadelphia, PA 19122, USA}
\affiliation{Institute for Computational Molecular Science, Temple University, Philadelphia, PA 19122, USA}

\date{\today}

\begin{abstract}
\par Accurate \textit{ab initio} theory of H-bond structure of liquid water requires high-level exchange correlation approximation from density functional theory. 
Based on the liquid structures modeled by \textit{ab initio} molecular dynamics by using maximally-localized Wannier functions as basis, we study the infrared spectrum of water within the canonical ensemble. 
In particular, we employ both Perdew-Burke-Ernzerhof (PBE) functional within generalized gradient approximation (GGA) and the state-of-art meta-GGA level approximation provided by the strongly constrained and appropriately normed (SCAN) functional. 
We demonstrate that the SCAN functional improves not only the water structure but also the theoretical infrared spectrum of water. 
Our analyses show that the improvement in the stretching and bending bands can be mainly attributed to the better descriptions of directional H-bonding and the covalency at the inter- and intra-molecular levels, respectively. 
On the other hand, the better agreements in libration and hindered translation bands are due to the improved dynamics of the H-bond network enabled by less structured liquid towards the experimental direction. 
The predicted spectrum by SCAN shows a much better agreement with experimental data compared to the conventionally widely adopted PBE functional at the GGA level.
 \end{abstract}

\maketitle

\section{INTRODUCTION}

\par As one of the most important materials on earth, water has an enormous impact on life. 
The unique functionalities of water lie in its liquid structure as depicted by the H-bond network. 
In this nearly tetrahedral structure, water molecules on neighboring sites are attracted by highly directional H-bonds, which constantly break and reform under thermal fluctuations at ambient conditions. 
Not surprisingly, the fundamental understanding of liquid water properties is at the center of scientific interests \cite{swartz_ab_2013,thamer_ultrafast_2015,kim_maxima_2017,Debenedetti_chemical_2017,chen_hydroxide_2018}.

\par Advanced experimental methods, such as scattering \cite{soper_radial_2013,skinner_benchmark_2013} and spectroscopy \cite{fransson_x-ray_2016} techniques, have been developed and applied to detect the nature of H-bond in water. 
In particular, the infrared (IR) spectroscopy provides a unique probe, in which both molecular configuration and its dynamics dielectric response can be inferred from the measured spectra \cite{brubach_signatures_2005,auer_hydrogen_2007}. 
In the IR spectrum, four main spectral features have been identified in experiments. 
Located in the relatively higher frequency range, the stretching and the bending bands can be traced back to the molecular vibrations in water vapor, which are strongly modified due to the presence of H-bond network in condensed phase. 
On the other hand, the libration and the hindered translational bands with lower frequencies originate from the collective motion of water molecules in the H-bonded network. 
Therefore, such vibrations have no counterparts in a single molecule.  

\par $Ab$ $initio$ molecular dynamics (AIMD) simulation provides an ideal theoretical framework to study the IR spectra in water from first principles \cite{car_unified_1985}. 
In AIMD simulations, water structures can be modeled by AIMD trajectories at finite temperatures, 
in which the forces acting on nuclei are obtained by the electronic ground state determined by the density functional theory (DFT) \cite{hohenberg_inhomogeneous_1964,kohn_self-consistent_1965}. 
The direct calculation of IR spectrum is allowed by the advent of modern theory of polarization given by the Berry phase formulation for extended systems \cite{king-smith_theory_1993}. 
The detailed dynamics dipolar correlation and its dependence on the H-bond network were further revealed later \cite{sharma_intermolecular_2005}, 
The above were facilitated by a rigorous decomposition of overall polarization onto the electric dipoles belonging to individual water molecules based on maximally-localized Wannier functions (MLWFs) \cite{marzari_maximally_2012}. 

\par Despite the above progress, difficulties remain when the computed IR is compared with available experiments. 
It is recognized that the accuracy of the predicted water structure depends on the level of adopted exchange-correlation functional in DFT. 
For the simulations of water, the GGA \cite{perdew_generalized_1996} is widely applied. 
However, the GGA functional significantly overestimates the H-bonding strength as well as the polarizability of water, which is evidenced by the large red shift ($\sim$200 cm$^{-1}$) \cite{sharma_intermolecular_2005, chen_role_2008}  of the computed stretching band compared to the experimental data. 
The above deficiency is partially due to the inherited self-interaction error.
As a result, one electron state applies a spurious electrostatic interaction on itself \cite{perdew_self-interaction_1981}.
By mixing a fraction of exact exchange, recent simulations showed that the underestimated stretching band in IR spectra of water can be largely improved by the PBE0-based AIMD simulation \cite{zhang_first_2011}. 
However, the application of hybrid DFT based AIMD demands significantly increased computational cost. 
Due to this reason, the available AIMD trajectories were relatively short, and the statistics was limited. 
Moreover, all the conventional functional approximations at the GGA level lack the description of long-range van der Waals (vdW) interactions. 
The long-range vdW is the key physical factor behind the larger mass density of water than that of ice \cite{zhang_structural_2011,chen_ab_2017}. 
Even in the canonical (NVT) ensemble, where the density of water is fixed to the experimental value at ambient conditions, the vdW inclusive AIMD simulation was found to have large effects on the water structure, 
which can be seen by the enhanced population of water molecules in the interstitial region \cite{schmidt_isobaricisothermal_2009,hermann_first-principles_2017}. 
It is accepted that the long-range vdW interaction has relatively small effect on the directional H-bonding. 
However, water molecules in the interstitial region are expected to be weakly bonded or non-bonded. 
So far, it is unclear whether or not the increased fluctuation of water molecules in the interstitial region will affect the overall IR spectra in a nontrivial way. 
Clearly, an improved modeling of the water structure and its dynamics via the DFT approach is prerequisite to answer the above questions. 

\par To address the above issues, we compute the IR spectrum of water from a systematically improved modeling of liquid water by the SCAN \cite{sun_strongly_2015} meta-GGA functional. 
The IR spectrum computed from the conventional PBE-GGA \cite{perdew_generalized_1996} AIMD trajectory is also presented here for comparison. 
By satisfying all the seventeen known exact constraints on the semi-local functional, the recently developed SCAN functional presented a greatly improved description in both covalency in water cluster \cite{sun_accurate_2016} and H-bond network in liquid water and ice \cite{chen_ab_2017,zheng_structural_2018}. 
Consistently, our computed IR spectrum based on SCAN-AIMD shows a significant improvement over the entire spectral range compared to the spectrum generated from conventional PBE-GGA functional. 
The red frequency shift and overestimated amplitude of the stretching band obtained with the PBE-GGA functional are largely corrected by the improved directional H-bonding strength. 
By using the MLWFs as basis, the increased population of water molecules in the interstitial region  is found to be anticorrelated which partially contributes to the decreased spectral amplitude towards the measured IR spectrum. 
The better agreement of the bending band frequency can be mainly attributed to the improved description of covalency of water molecules. 
On the other hand, the blue shift of the libration modes towards the experimental direction is correlated to the improved prediction of dynamical property in water.

\section{METHODS}
\label{part2}
\par We computed the IR spectra of liquid water based on trajectories from Car-Parrinello molecular dynamics \cite{car_unified_1985} using the Quantum Espresso package \cite{Giannozzi_advanced_2017}. 
The PBE and SCAN exchange-correlation functionals were used.
Norm-conserving pseudopotentials \cite{hamann_optimized_2013} were adopted with an energy cutoff of 85 Ry. 
All simulations were carried out in a periodically cubic box with side lengths of 23.57 Bohr ($12.47$ \r{A}), and 64 D$_2$O molecules were included in the box. 
The time step was set to 2 a.u. ($\sim$0.048 fs), with electron mass as 100 a.u.. 
The temperature was set to 330 K and the Nos\'e-Hoover thermostat \cite{nose_unified_1984,hoover_canonical_1985} was adopted with the canonical ensemble. 
The 30 K increases to ambient temperature are applied to roughly mimic nuclear quantum effects on liquid structures, especially the oxygen-oxygen pair distribution function, and be consistent with previous studies\cite{Morrone_nuclear_2008,chen_ab_2017}.
All simulations were run for more than 50 ps.
In addition, we carried out ground-state DFT calculations for water monomer in a cubic box with cell lengths being 30 Bohr (15.88 \r{A}), using both SCAN and PBE with an energy cutoff of 250 Ry, to calculate the vibrational frequencies.

\par The IR spectra of liquid water are computed based on the IR absorption rate in terms of the total dipole moment.
The formula \cite{mcquarrie_statistical_2000} is derived using the Fermi's golden rule and Poynting vector with a classical approach \cite{bader_quantum_1994} applied, where we treat the quantum time correlation function classically:
\begin{equation} 
\label{eq_ir}
\alpha(\omega)n(\omega) = \frac{2\pi\beta\omega^2}{3cV} \int_{-\infty}^{+\infty} dt{e}^{-i\omega t} \langle M(0)M(t) \rangle
\end{equation}
where $\alpha(\omega)$ is the light absorption coefficient per unit depth as a funcition of the frequency $\omega$, 
$n(\omega)$ is the refractive index, $\beta = (k_BT)^{-1}$ with $k_B$ and $T$ being the Boltzmann factor and temperature, respectively. 
The total dipole moment $M$ in the simulation cell is computed via the formula: $\sum_{i=1}^{n}\mu_i$,
where the molecular dipole moment of the $i$th water molecule $\mu_i$ can be calculated via the position of nuclei and corresponding Wannier centers.
Furthermore, we adopted a gaussian window \cite{harris_use_1978} in the form of $e^{-t^2/2\alpha^2}$ with $\alpha=0.5$ ps, when the discrete Fourier transform in Eq. \ref{eq_ir} is performed.

\section{RESULTS and DISCUSSIONS}
\label{sis}

\begin{figure}[h]
\centering
\includegraphics[width=\columnwidth]{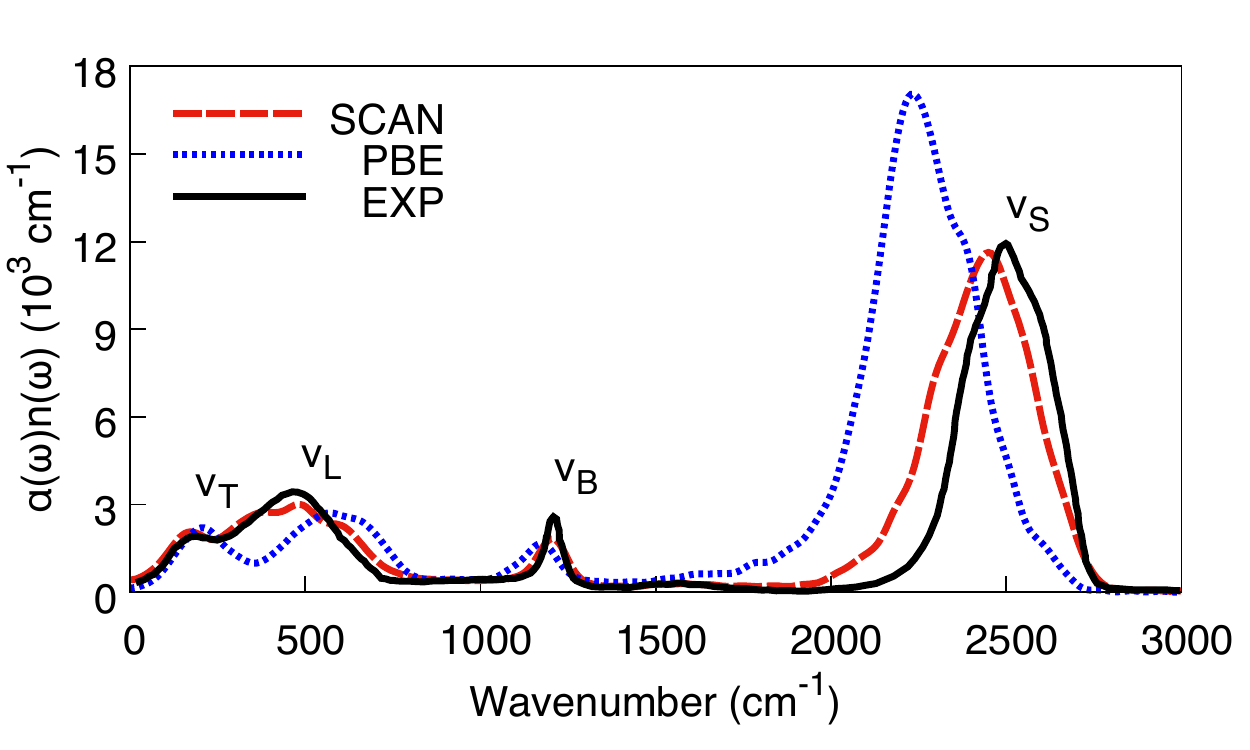}
\caption{(Color online) IR spectra of liquid water obtained from AIMD simulations in the NVT ensembles. The experimental data are from Refs. \onlinecite{max_isotope_2009} at room temperature. Labels $\nu_T$,  $\nu_L$,  $\nu_B$, and $\nu_S$ correspond to translation, libration, DOD bending, and OD stretching peaks, respectively. }
\label{fig1}
\end{figure}

\par As shown in Fig. \ref{fig1}, we present the theoretical IR spectra generated by SCAN-AIMD and PBE-AIMD trajectories. 
For comparison, the experimental spectrum is also shown. 
Four main dipole correlated vibrational bands can be identified in Fig. \ref{fig1}, 
whose characteristics belong to the hindered translation (T), libration (L), deuterium-oxygen-deuterium (DOD) bending (B), and the oxygen-deuterium (OD) stretching (S) modes with increasing frequencies. 
The IR spectrum computed from SCAN-AIMD shows a significantly improved accuracy comparable to that computed with the hybrid DFT functional \cite{zhang_first_2011}, 
in which the better agreement with the experiment can be seen in both the spectral positions and spectral shape. 
The peaks of the above four bands, calculated by SCAN-AIMD, are centered at $\nu_T$=172 (186) cm$^{-1}$, $\nu_L$=483 (486) cm$^{-1}$, $\nu_B$=1207 (1209) cm$^{-1}$, $\nu_S$=2448 (2498) cm$^{-1}$. 
In comparison, the same bands obtained from PBE-AIMD are predicted to be centered at $\nu_T$=207 (186) cm$^{-1}$, $\nu_L$=572 (486) cm$^{-1}$, $\nu_B$=1174 (1209) cm$^{-1}$, and $\nu_S$=2233 (2498) cm$^{-1}$ respectively.
In the above, the values presented in the parenthesis are taken from experiment \cite{max_isotope_2009}. 
In particular, the improvement by SCAN in the positions of the liberation and stretching modes is significant, 
which can be seen by the frequency shift of $\sim$100 and $\sim$200 cm$^{-1}$ respectively.
At the same time, the spectral shape is also largely improved towards the experimental direction as clearly evidenced by the reduced intensity in the stretching band over 30\%, 
which is severely overestimated at the PBE level.


\begin{table}[t]
\centering
\caption{Water monomer (D$_2$O) vibration frequencies in cm$^{-1}$. (Subscripts $b$,$ss$, and $as$ stand for bending, symmetric stretching, and anti-symmetric stretching modes, respectively.)}
\label{vf}
\begin{ruledtabular}
\begin{tabular}{p{0.23\columnwidth}<{\centering}p{0.23\columnwidth}<{\centering}p{0.23\columnwidth}<{\centering}p{0.23\columnwidth}<{\centering}}
Methods&$\nu_b $&$\nu_{ss}$&$\nu_{as}$\\
\hline
SCAN&1193&2710&2832\\
PBE&1162&2660&2770\\
EXP \cite{benedict_rotationvibration_1956}&1206&2784&2889\\
\end{tabular}
\end{ruledtabular}
\end{table}

\par The OD stretching band originates from the molecular vibrations in water vapor (gas) phase. 
Based on the C$_{2v}$ point group symmetry of water monomer, these vibrations can be further categorized as symmetric and anti-symmetric eigen modes with a slightly higher frequency of the later in experiments. 
Under the development of stretching modes, the covalent bond is elongated with a tendency towards dissociation, which also directly modifies the electric dipole of water molecules. 
Not surprisingly, the stretching band in the IR spectra not only locates at the highest frequency range but also has the most prominent spectral intensity among the four IR spectral features. 
At the PBE level, the frequencies of both symmetric and anti-symmetric modes in gas phase are underestimated over  $\sim$4\% compared to experiment as shown in Table \ref{vf}, which is consistent with the previous calculations \cite{zhang_first_2011}. 
Compared to PBE, the SCAN functional predicts better stretching frequencies in water gas phase with improvement of $\sim$2\% towards experiment as shown in Table \ref{vf}. 
The above can be attributed to the improved prediction of covalency in water monomer. 
Indeed, the better agreement with experiments by SCAN functional in terms of bond angle, bong length, 
and binding energies in single water molecule and water clusters in gas phase have been widely recognized recently \cite{sun_accurate_2016,zheng_structural_2018}.  

\par
\begin{figure}[t]
\centering
\includegraphics[width=0.7\columnwidth]{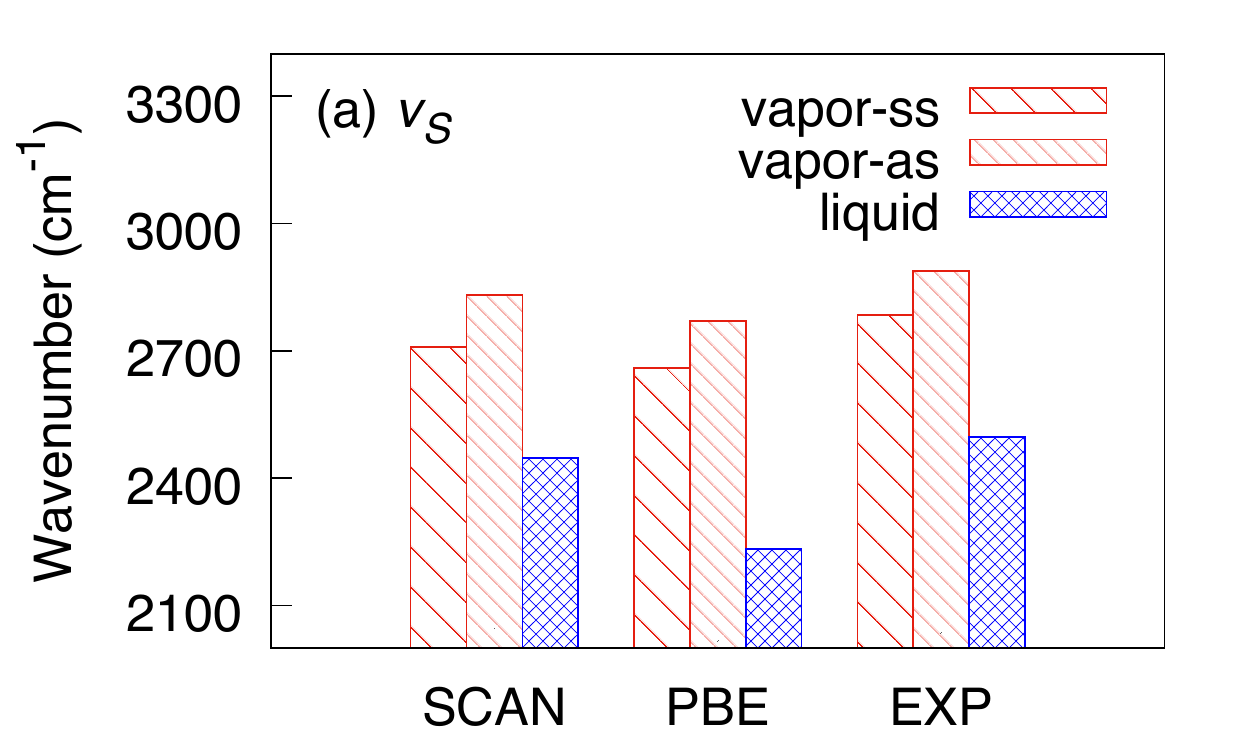}
\includegraphics[width=0.7\columnwidth]{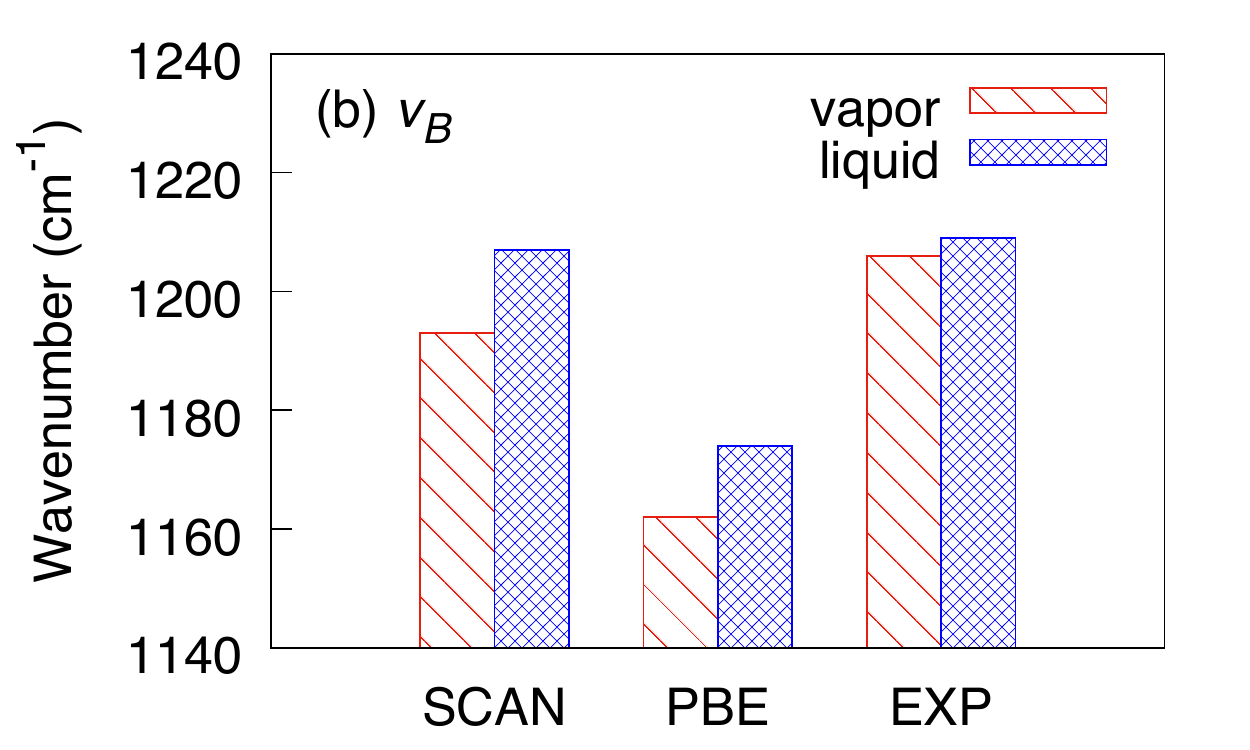}
\caption{(Color online) (a) Stretching frequency $\nu_S$ and (b) bending frequency $\nu_B$ of single water molecule (vapor) and liquid water as compared using SCAN, PBE, and experiment data (EXP) of vapor \cite{benedict_rotationvibration_1956} and liquid water \cite{max_isotope_2009}. (Note that symmetric stretching frequency $\nu_{ss}$ and anti-symmetric stretching frequency $\nu_{as}$ are listed separately for water vapor.)}
\label{vapor}
\end{figure}

 In the liquid phase, the difference between symmetric and anti-symmetric modes is smeared by the disordered environment and the stretching band is formed. 
Moreover, the H-bond, represented by the weak attraction of a proton to the oxygen lone pair of the neighboring molecules, 
also facilitates the elongation of OH bonds as shown by the red shift of the stretching vibration by $\sim$300 cm$^{-1}$ in liquid compared to the corresponding vibration in vapor phase as measured experimentally in Fig. \ref{vapor} (a). 
However, due to the significant overestimation of the H-bond strength by PBE functional, the stretching frequency of liquid water is predicted to be $\sim$500 cm$^{-1}$ lower than that of water in vapor phase. 
The artificially strengthened H-bond by PBE also leads to the overestimated dipole-dipole correlation in the H-bond network as evidenced by the much greater IR spectral intensity than the experimental spectrum in Fig. \ref{fig1}. 
In sharp contrast, the IR spectrum computed by SCAN functional shows a significantly better agreement in the above aspects. 
We attribute this improvement to the more accurate H-bond structure as well as dynamical correlation described by SCAN functional.

\begin{figure*}[!t]
\centering
\includegraphics[width=0.99\columnwidth]{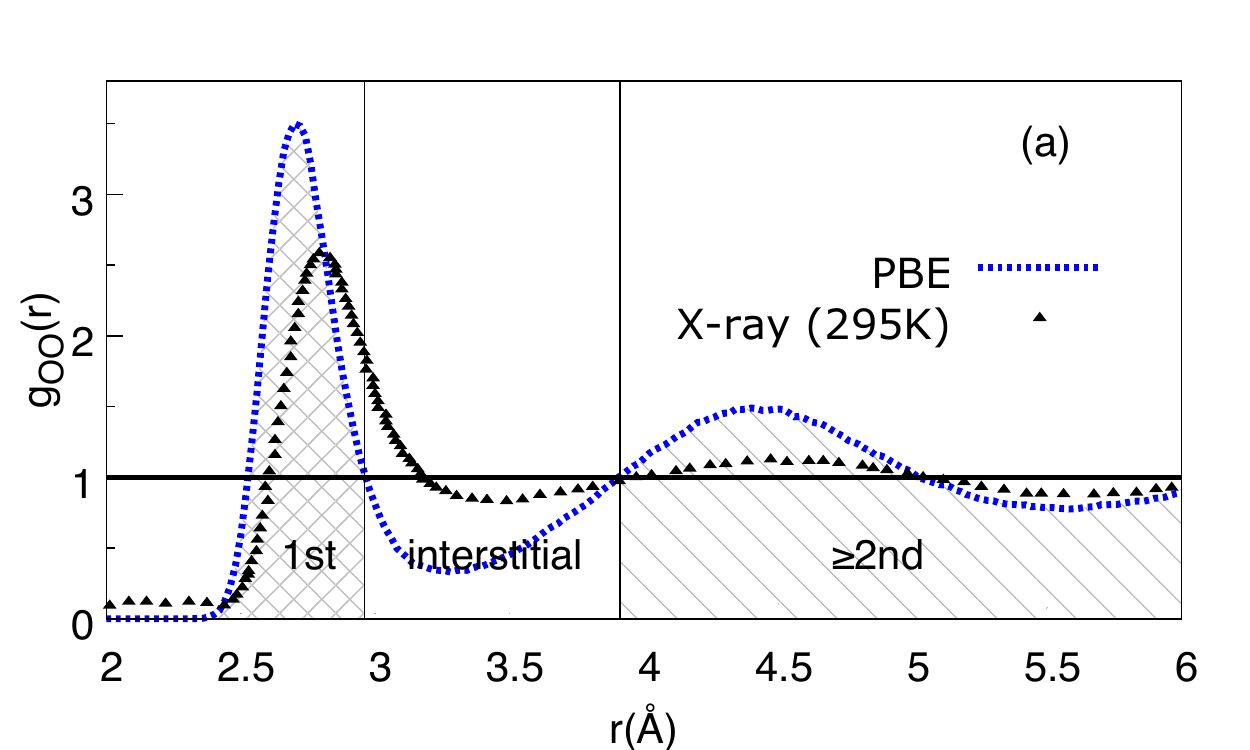}
\includegraphics[width=0.99\columnwidth]{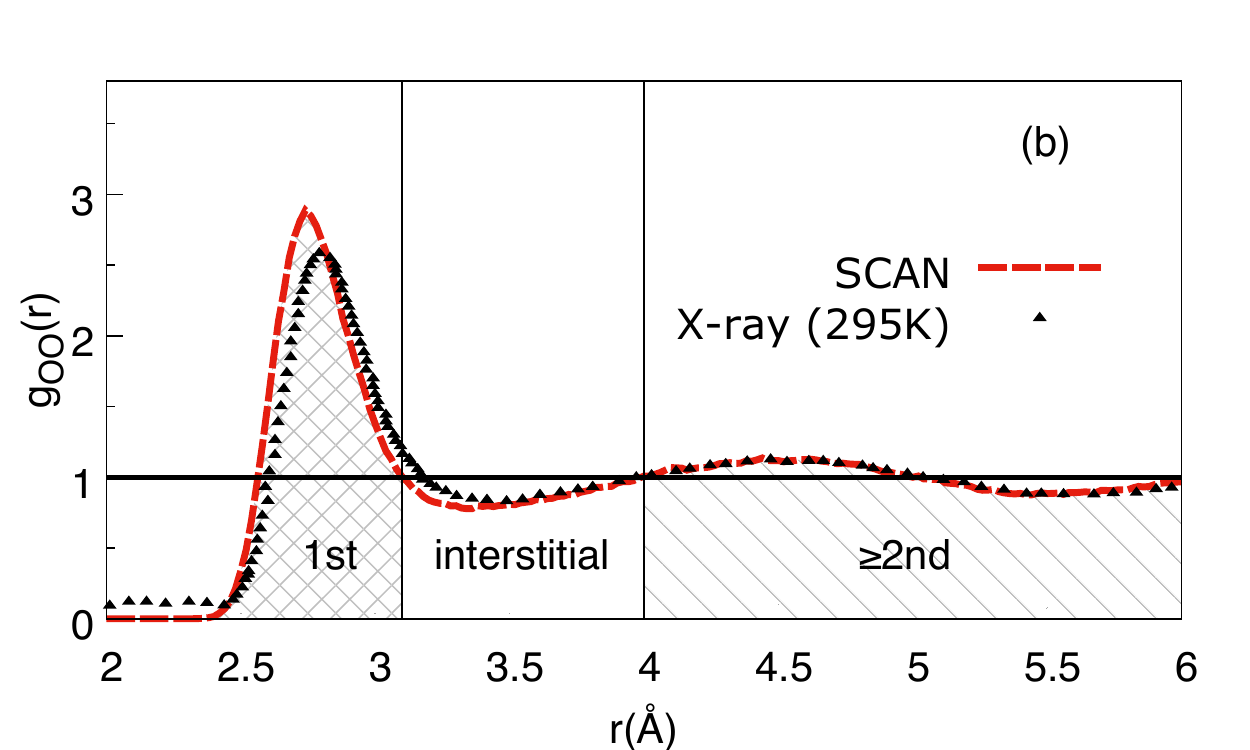}
\includegraphics[width=0.99\columnwidth]{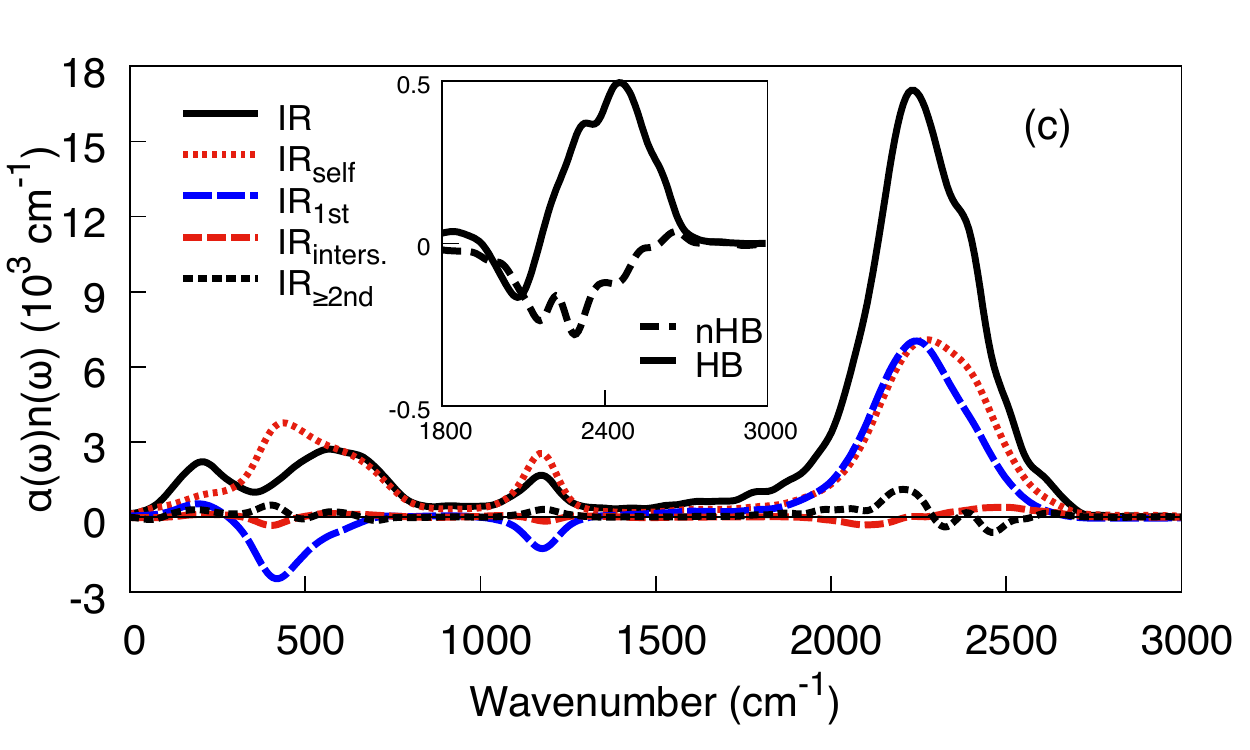}
\includegraphics[width=0.99\columnwidth]{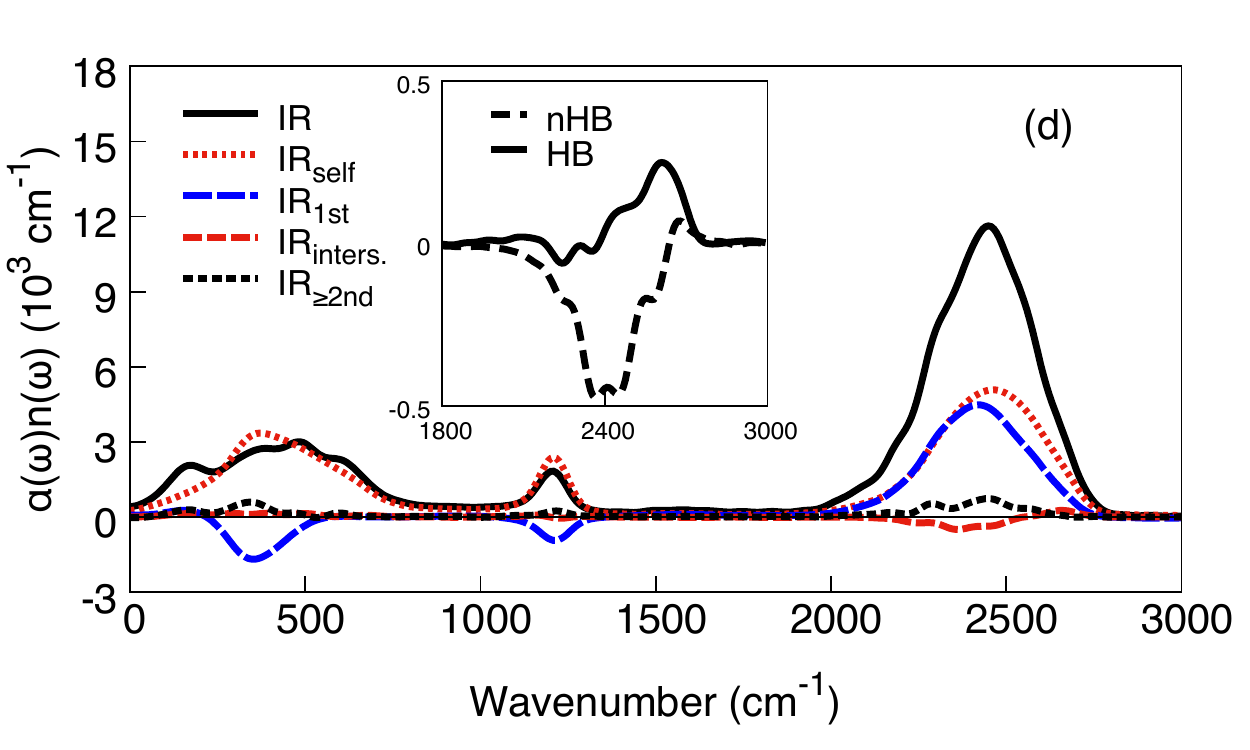}
\caption{(Color online) Oxygen-oxygen pair distribution functions and the corresponding spacial decompositions by (a) PBE and (b) SCAN functionals compared with X-ray diffraction experiment \cite{skinner_benchmark_2013}. Intramolecular ${\rm IR_{self}}$ (red dotted lines) and spacial intermolecular ${\rm IR_{1st}}$ ${\rm IR_{inters.}}$, and ${\rm IR_{\geq 2nd}}$ (colored dashed lines) contributions to the calculated IR spectra (black solid lines) using (c) PBE and (d) SCAN in the NVT ensemble. Decompositions of ${\rm IR_{inters.}}$ into non H-bonded (dashed lines) and H-bonded (solid lines) water contributions are shown as insets in (c) and (d).}
\label{fig3}
\end{figure*}

\par In order to elucidate the spectral signatures of liquid structure, we further decompose the overall spectrum into contributions from different ranges in the H-bond network. 
Based on the method introduced by Chen et al. \cite{chen_role_2008}, 
the total dipole correlation function in the temporal domain of Eq.(\ref{eq_ir}) can be rewritten as the sum of a correlation function of any pair of water molecules as $\langle M(0)M(t)\rangle(r_1,r_2)=\langle\sum_{i,j}\mu_i(0)\mu_j(t)\rangle$. 
In the above, the electronic and ionic contributions of dipole moment $\mu_i$ and $\mu_j$ of water molecules at any time $t$ are rigorously determined by the Wannier centers and the ionic coordinates respectively, 
both of which are generated on the fly in the AIMD simulation. 
Therefore, the spectrum can be divided into intramolecular  ${\rm IR_{self}}$ ($i=j$) and intermolecular ($i\neq j$) parts \cite{sharma_intermolecular_2005}. 
In current work, we further divide the intermolecular part into spectral contributions from the first coordination shell ${\rm IR_{1st}}$, water in the interstitial region ${\rm IR_{inters.}}$, and water from the second shell and beyond ${\rm IR_{\geq 2nd}}$\footnote{In order to sample the angular distribution uniformly while calculating the IR spectra of second shell and beyond region. We apply an upper limit of the correlated distance as half of the box length.}, respectively based on the distance $r_{ij}$ between the molecule pairs as illustrated in Fig. \ref{fig3} (a) and (b). 
The criterion for interstitial region is chosen to be $3.10$ \r{A} $\leq r_{ij} < 4.00$ \r{A} in SCAN and $2.96$ \r{A} $\leq r_{ij} < 3.91$ \r{A} in PBE,
where the oxygen-oxygen pair distribution function ${\rm g_{OO}}(r)$ is less than 1 between the first and second peaks.
In the decomposition, water molecules are considered as H-bonded if the O-O distance is less than 3.5 \r{A} and the OOD angle is less than 30$^{\circ}$\cite{luzar_hydrogen-bond_1996}.
The resulting spectral decompositions and the corresponding ${\rm g_{OO}}(r)$ are shown in Fig. \ref{fig3} for both PBE and SCAN functionals. 

\par Dynamically, the stretching mode not only modifies the electric dipole of the water molecule under vibration, 
but also strongly polarizes the surrounding water molecules via the H-bonds. 
Therefore, the stretching band of IR spectrum is dominated by both ${\rm IR_{self}}$  and ${\rm IR_{1st}}$ contributions. 
Such effect has been well captured by both SCAN and PBE functionals 
which can be seen by the large and comparable intensities from these two decompositions. 
In the above, the ${\rm IR_{self}}$ has a slightly higher frequency than that of ${\rm IR_{1st}}$. 
It is not surprising since the ${\rm IR_{1st}}$ is more sensible to the H-bond network,
while the ${\rm IR_{self}}$ is instead more affected by the intrinsic vibration of water monomer. 
However, the IR spectrum from PBE functional suffers from clear drawbacks. 
Compared to the experimental measurement, 
the ${\rm g_{OO}}(r)$ is significantly over-structured with a shorter first peak position than experimental measurement, 
indicating the overestimated H-bond strength, as shown in Fig. \ref{fig3} (a). 
As a result, the stretching band of IR spectrum computed at the PBE level is centered at a much lower frequency of 2233 cm$^{-1}$ with a much higher intensity in comparison to experimental measurements.
It is consistent with the fact that PBE predicts the proton to be more delocalized and easier to be donated to the neighboring water molecules than it should be. 
Such a severely overestimated directional H-bond strength is largely improved by SCAN functional, 
which can be seen by the less structured coordination shell and increased distance of the first peak in ${\rm g_{OO}}(r)$, as shown in Fig. \ref{fig3} (b). 
The softer H-bond network towards the experimental direction also generates a less polarizable liquid water, 
as indicated by the predicted average electric dipole by SCAN (2.95 $\pm$ 0.28 D) being much closer to the experimental reference (2.9 $\pm$ 0.6 D) \cite{badyal_electron_2000} than that of PBE (3.22 $\pm$ 0.29 D).  
As expected, the stretching mode in the less polarizable medium becomes harder 
and shifts the stretching band center to a higher frequency at 2448 cm$^{-1}$ 
as well as reduces intensities mainly from both ${\rm IR_{self}}$  and ${\rm IR_{1st}}$. 

\par The dynamical polar correlation from stretching vibration decays rather rapidly for water molecules separated by a distance further away from the first coordination shell, 
which are qualitatively similar for both PBE and SCAN predictions as shown in Figs. \ref{fig3} (c)(d). 
Indeed, it is consistent with the weaker structural correlations in the radial distribution ${\rm g_{OO}}(r)$ beyond the first peak in Figs. \ref{fig3} (a)(b). 
However, the drawbacks can be still identified due to the over-structured liquid water by PBE functional. 
On the one hand, the spectral contribution from water in the second shell and beyond ${\rm IR_{\geq 2nd}}$ is predicted to have a higher intensity in PBE than that from SCAN. 
On the other hand, due to the artificially strengthened H-bond strength, most water molecules in the interstitial region are still H-bonded to the central molecule under stretching motion, 
giving rise to the positively spectral intensity in the insert of Fig. \ref{fig3} (c). 
However, the opposite trend is observed in the prediction by SCAN functional, 
in which the water molecules in the interstitial region are mostly non-bounded with the observed anti-correlation as seen by the negative spectral intensity in the insert of Fig. \ref{fig3} (d). 
It should be noted that the increasingly populated water molecules in the interstitial region should be attributed to the intermediate range vdW interactions captured by SCAN function, 
which is found to be the key physical effect in predicting a higher mass density of water than that of ice \cite{chen_ab_2017}. 
With the weaker dipole correlations from water molecules beyond the first coordination shell that correctly predicted by SCAN functional, 
the overestimated IR spectral intensity by PBE is further reduced towards the experimental direction.

\par Moreover, a shoulder around $\sim$2300 cm$^{-1}$ on the left-hand side of the peak is identified in the stretching band in experimental spectra of water. 
According to recent theoretical analyses\cite{kananenka_fermi_2018,hunter_disentangling_2018}, this shoulder feature is attributed to the so-called Fermi resonance which describes the spectral enhancement due to the intramolecular vibrational coupling between the OD stretching mode and the overtone of bending mode. 
At both PBE and SCAN levels of studies, the shoulder appears roughly at the same position around 2300 cm$^{-1}$.
This is not surprising since the bending overtone in this frequency range carries twice the frequency of water bending motion, and both PBE and SCAN yield a reasonably accurate bending frequency around 1200 cm$^{-1}$. 
Nevertheless, the relative position of the Fermi resonance shoulder feature is misplaced, which shows a blue shift appearing on the right-hand side of the stretching peak in the IR spectroscopy at PBE level in Fig. \ref{fig1}.
In sharp contrast, as the SCAN functional accurately predicted the OD stretching frequencies, the shoulder emerges clearly as shown in Fig. \ref{fig1}. 
At the same time, more pronounced enhancement of IR spectral amplitudes is observed for ${\rm IR_{self}}$ and ${\rm IR_{1st}}$ in the above frequency region for the spectral by SCAN. 
The above is exactly consistent with intramolecular vibrational coupling nature of the Fermi resonance\cite{kananenka_fermi_2018,hunter_disentangling_2018}.

\par Among the four spectral features, the bending band has the second largest frequency centered at around 1200 cm$^{-1}$, 
which originates from the bending mode of water monomer. 
Under the bending mode, the DOD bonding angle is modulated, 
which in turn changes the electric dipole of the molecule, 
but with a much weaker coupling strength than stretching mode. 
Not surprisingly, the resulting spectral intensity is also smaller than the stretching band. 
In vapor phase, the frequency of the bending mode is underestimated by $\sim$4\% at PBE level, 
while the calculation by SCAN functional gives rise to a much more accurate value with an error of $\sim$1\% compared to experiment at 1206 cm$^{-1}$ in Table \ref{vf} and Fig. \ref{vapor} (b). 
Clearly, such a significant improvement should be again attributed to the better description of the covalency by SCAN. 
Indeed, the bonding angle and bonding length of water monomer are respectively found to be 104.4$^{\circ}$ (104.5$^{\circ}$) and 0.961 \r{A} (0.957 \r{A}) by SCAN, 
showing a large improvement over the PBE functional of 104.2$^{\circ}$ and 0.970 \r{A} as compared to the experimental values in the parenthesis \cite{sun_accurate_2016}.  
 
\par In condensed phase, the bending band instead of discrete levels is formed by the disordered liquid structure. 
According to the spectral decompositions in Figs. \ref{fig3} (c)(d), 
the bending band is mainly contributed by the intramolecular contribution and the first coordination shell. 
In particular, the water molecules in the first coordination shell provide a large negative spectral intensity resulting from its anti-correlation in nature. 
It is not surprising since the protons under bending mode move along the normal direction of H-bond. 
Therefore, the development of bending vibration needs to overcome the energy to break the H-bond and generate out-of-phase dynamical dipole correlation \cite{chen_role_2008}. 
As a result, unlike the stretch band, the presence of H-bond network in liquid water impedes the bending mode resulting in a slightly increased frequency 
relatively to that in water vapor in experiment. 
Such a feature has been qualitatively predicted in both PBE and SCAN functionals. 
Clearly, the frequency difference is still overestimated by roughly the similar magnitude for both functionals. 
The deficiency is likely due to the self-interaction error inherited in both GGA and meta-GGA functionals resulting in the delocalized protons that are more easily to be donated to neighboring oxygen atom.


\begin{figure}[t]
\centering
\includegraphics[width=\columnwidth]{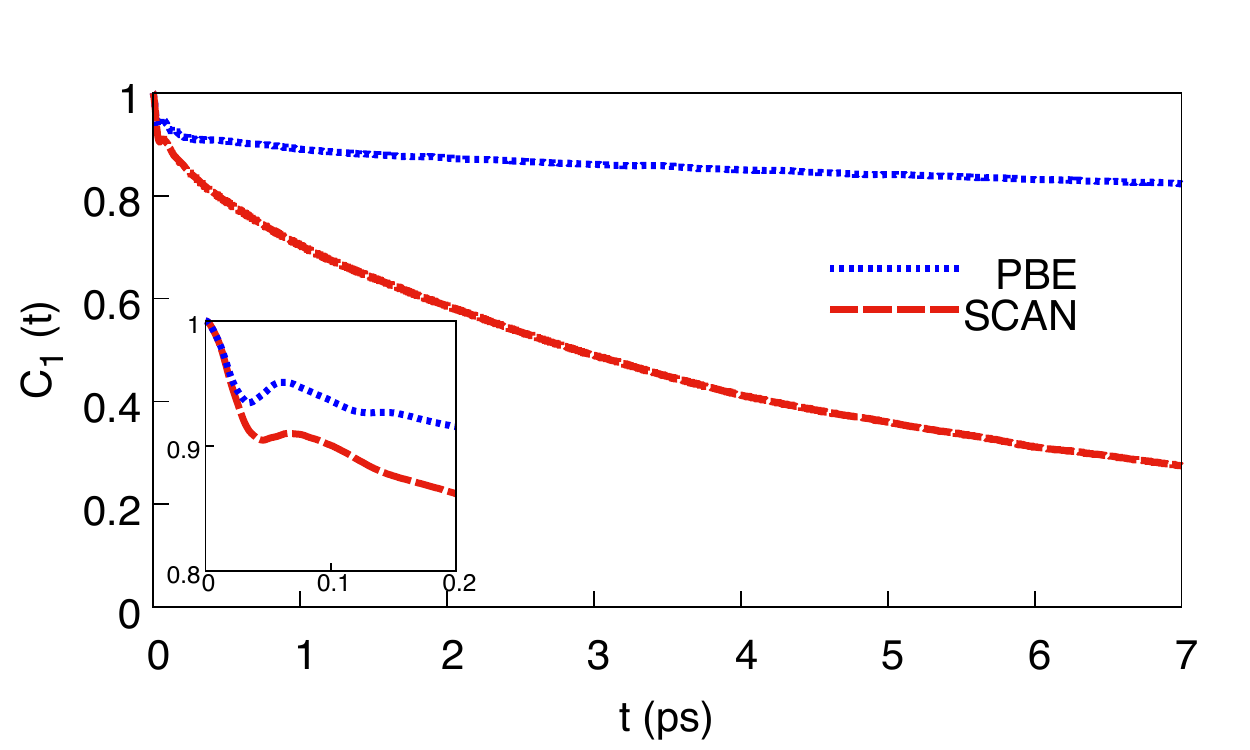}
\caption{(Color online) First order rotational correlation functions of liquid water obtained from AIMD simulations with SCAN (red dashed line) and PBE (blue dotted line) in the NVT ensembles. The inset shows enlarged rotational correlation functions in the first 0.2 ps.}
\label{fig5}
\end{figure}

\par In the far-infrared region in Fig. \ref{fig1}, 
the two spectral features in a rather broad distribution are associated with the collective vibrations on the H-bond network, 
which emerge in the condensed phases such as ice and water. 
The feature at the higher frequency branch around $\sim$500 cm$^{-1}$ is attributed to the libration mode. 
Depicted by water libration, the water molecules undergo hindered rotational motions restricted by the presence of H-bond network. 
Therefore, such spectral signal is absent in the water vapor since it is free to rotate an isolated molecule. 
Like the bending motion, water molecules under libration also needs to overcome the energy barrier by breaking H-bonds. 
Consistently, the spectral decomposition from intramolecular and first coordination contributions are dominant, 
in which the latter is characterized by the anti-correlation with negative spectral intensities as shown in Figs. \ref{fig3} (c)(d). 
Because of the over-structured H-bond network, 
the frequency of the libration motion is exaggerated $\sim$20\% by PBE functional, centered at 572 cm$^{-1}$. 
The less structured liquid water modeled by the SCAN functional eases the water libration at 483 cm$^{-1}$, 
matching accurately the experimental data at 486 cm$^{-1}$. 
Because the direction of water polarity aligns along HOH angular bisector, 
the hindered water rotation is also highly associated with dynamical properties such as the rotational diffusion constant and rotational relaxation time which is accessible to experiments. 
The relaxation time $\tau_n$ is obtained by fitting with $e^{-t/\tau}$ to the $n$th order rotational correlation functions, 
which are defined by the time correlation function \unexpanded{$C_n(t) = \left \langle P_n[\boldsymbol{\hat{u}}(0)\boldsymbol{\hat{u}}(t)] \right \rangle$}. 
Here $P_n$ is the $n$th Legendre polynomial and $\boldsymbol{\hat{u}}$ is the in plane unit vector along the symmetry axis of a water molecule. 
To this end, we compute the first order rotational correlation function $C_1(t)$ and the corresponding relaxation time $\tau_1$ based on trajectories from both PBE and SCAN functionals as shown in Fig. \ref{fig5}. 
It should be noted that features at the beginning of rotational correlation function around 0.066 and 0.062 ps, magnified in the inserts of Fig. \ref{fig5},
are corresponding to the libration motion that have been observed in the frequency domain in Fig. \ref{fig3}. 
As time passes, the rotational correlation decays in both trajectories giving rise to the relaxation time $\tau_1$. 
Under the Debye's model, 
the relation between relaxation time $\tau_n$ and the rotational diffusion constant $D_R$ takes the form of $\tau_n = 1/[n(n+1)D_R]$. 
In the above, the rotational diffusion constant $D_R$ is given by the inverse of $\tau_1$, which is measured to be 2.0-7.5 ps \cite{laage_molecular_2008}. 
Clearly, the over-structured water model by PBE functional hampers the rotation motion and 
predicts an unphysically long relaxation time $\tau_1$=$20.91(\pm0.22$) ps with the corresponding small rotational diffusion constant $D_R$=0.047 ps$^{-1}$.  
In sharp contrast, the less structured water modeled by SCAN functional largely facilitates the water rotation and 
gives rise to much more reasonable values of $\tau_1$=$4.08 (\pm 0.02$) ps and $D_R$=0.24 ps$^{-1}$, respectively.
The result consists with previous studies on diffusion coefficient\cite{zheng_structural_2018}.
As a conclusion, the modified libration peak from the SCAN functional indicates an improved description of the diffusion related dynamics in liquid water.

\par In the gas phase, the water molecule is free to undergo a translation motion in space. 
In liquid, water molecules however undergo hindered translation constrained by the H-bond network, 
which is depicted by the lower branch feature of the far-infrared band centered at $\sim$200 cm$^{-1}$ in Fig. \ref{fig1}. 
Similar to the stretching mode, the spectral decomposition in Fig. \ref{fig3} shows that 
the IR signals are mainly contributed by the dynamical correlation from intramolecular contribution and molecules in the first coordination shell, 
which is consistent with the first-principles studies of Chen et al \cite{chen_role_2008}. 
Unlike the stretching mode that is directly coupled to the electric dipole, 
the dynamical correlation here is generated by the induced dipole-dipole interaction under the hindered translation motion \cite{guillot_molecular_1991}. 
Therefore, the intensities of hindered translation are much weaker than those in the stretching band as shown in Fig. \ref{fig3}. 
Not surprisingly, the artificially overstructured liquid water predicted by PBE functional overestimates the energy barrier cost of hindered translation, 
which is evidenced by an overestimated peak position at 207 cm$^{-1}$, compared to the experimental value of 186 cm$^{-1}$. 
On the other hand, the less structured liquid water model by SCAN functional facilitates the water translation with a slightly underestimated value of 172 cm$^{-1}$.

\par It is well accepted that the liquid water structure under PBE prediction is ice-like 
which is not only over structured but also sluggish that barely diffuses in space. 
As a result, the overall shape of the libration and hindered translation band is also ice like under the PBE prediction as shown in Fig. \ref{fig1}. 
In the above, the distinction between the two modes is exaggerated by the rather deep minimum 
with a frequency gap around 370 cm$^{-1}$. 
Indeed, in the crystalline ice, the libration and hindered translation are two distinct spectral bands, 
which are separated completely with frequencies of 640 cm$^{-1}$ and 222 cm$^{-1}$ 
in experiment \cite{eisenberg_structure_2005}, respectively. 
Strikingly, the above ice-like error is mostly corrected in the spectrum modeled by SCAN functional. 
For a liquid water structure that is more softened and disordered towards the experimental direction, 
the distinction between libration and hindered translation is also smeared as shown in Fig. \ref{fig1}.

\section{CONCLUSION}

\par In conclusion, we have performed careful and comparative first-principles molecular dynamics studies on the IR spectra of liquid water obtained by SCAN meta-GGA and PBE GGA functionals.
Our results showed the SCAN meta-GGA functional provides significant improvements on all four peaks in the IR spectra, compared to those obtained by PBE. 
Our analysis demonstrate that the SCAN functional models water more precisely on both molecular and liquid levels, and gives better descriptions of electronic structure, range dependent correlations, and dynamical properties. 
On one hand, SCAN reaches the accuracy level of the IR spectra obtained by PBE0 hybrid functional \cite{zhang_first_2011}, with a relatively low computational cost.
On the other hand, one may expect that the hybrid functional SCAN0 mitigates the self-interaction error in DFT, 
and brings the calculated IR spectra closer to the experiments.
Furthermore, recent studies \cite{Ceriotti_nuclear_2013,sun_electron-hole_2018,hunter_disentangling_2018} suggested that nuclear quantum effects due to light protons play a crucial role in obtaining accurate liquid water properties.
The role played by quantum nuclei awaits the future investigation which is likely to broaden the spectral features and slightly reduce the frequencies of stretching band due to the delocalized protons.

\begin{acknowledgments}
\par This work was supported as part of the Center for the Computational Design of Functional Layered Materials, an Energy Frontier Research Center funded by the U.S. Department of Energy, Office of Science, Basic Energy Sciences under Award No. DE-SC0012575. 
X. Wu is partially supported by  National Science Foundation through Awards No. DMR-1552287.
The computational work used resources of the National Energy Research Scientific Computing Center (NERSC), a U.S. Department of Energy Office of Science User Facility operated under Contract No. DE-AC02-05CH11231. And this research includes calculations carried out on Temple University's HPC resources and thus was supported in part by the National Science Foundation through major research instrumentation grant number 1625061 and by the US Army Research Laboratory under contract number W911NF-16-2-0189.
\end{acknowledgments}

\bibliography{Infrared_Spectra}

\end{document}